\documentclass[useAMS,usenatbib,usegraphicx,fleqn]{mn2e}
\usepackage{amsmath}
 \usepackage{times}
 
\title[Earthshine observations at high spectral resolution]
  {Earthshine observations at high spectral resolution: \\ Exploring and detecting metal lines in the Earth's upper atmosphere}
\author[B.Gonz\'alez-Merino et al. ]
 {B. Gonz\'alez-Merino,$^{1,2}$, E. Pall\'e,$^{1,2}$, F. Motalebi,$^{3,4}$, P. Monta\~n\'es-Rodr\'iguez,$^{1,2}$,
\newauthor 
and M. Kissler-Patig$^{4,5}$ \\
$^1$ Instituto de Astrof\'isica de Canarias, C/ v\'ia L\'actea, s/n, 38205 La Laguna, Tenerife, Spain\thanks{E-mail: bgmerino@iac.es}\\
$^2$ Dpto. de Astrof\'isica, Universidad de La Laguna, 38206 La Laguna, Tenerife, Spain \\
$^3$ Observatoire Astronomique de l'Université de Genève, Chemin des Maillettes 51, 1290 Sauverny, Switzerland \\
$^4$ European Southern Observatory, Karl-Schwarzschild-Strasse 2, 85748 Garching bei M\"{u}nchen, Germany \\
$^5$ Gemini Observatory, 670 N. A'ohoku Place, Hilo, Hawaii, 96720, USA}

\begin{document}

\date{}

\pagerange{\pageref{firstpage}--\pageref{lastpage}} \pubyear{2002}

\maketitle

\label{firstpage}

\begin{abstract}
Observations of the Earth as a planet using the earthshine technique (i.e. looking at the light reflected from the darkside of the Moon), have been used for climate and astrobiology studies. They provide information about the planetary albedo, a fundamental parameter of the Earth's energy balance. Here we present for the first time, observations of the earthshine taken at high spectral resolution. The high spectral resolution was chosen in order to investigate the possibility of detecting metallic layers in the Earth's atmosphere of geological or meteoritic origin. The SARG echelle spectrograph at the Telescopio Nazionale Galileo in La Palma was used to acquire the earthshine data. Observations were carried out on several nights in February 2011, with the spectral resolution set at 29,000, covering a spectral range from the near-ultraviolet (360 nm) to near-infrared (1011.9 nm). While we find evidence for the detection of a Na layer in the earthshine, other atomic species are not detected, perhaps due to the low signal-to-noise ratio of the observations and the difficult telluric corrections. The Na layer is found to vary between observation dates, which we speculate is due to physical variations in mesospheric Na concentrations. 
\end{abstract}


\section{Introduction}
Exoplanets have been one of the major areas of interests in astrophysics over the last few years. Since the first exoplanet detections \citep{Wolszczan92,Mayor95}, the number of discoveries has grown at an ever increasing rate. Although the majority of confirmed exoplanets are hot-Jupiter types, in recent years the first super-Earths (rocky bodies several times the size of the Earth) have already been identified (e.g. \citet{Rivera05}), and more recently, the discovery of the first Earth-like planet located in the habitable zone of the host star has been reported \citep{Borucki12}. However, the detection of small rocky planets is only the first step; the characterization of their atmospheres and even the possibility of detecting the fingerprints of life will be next. An understanding of the instrumental challenges involved in these atmospheric characterisation studies \citep{Lammer09} requires experience in the interpretation of planetary spectra. In this sense, observations of the Earth as a distant planet, which can be obtained using earthshine observations, will be a very useful template.

\begin{figure}
\centering
\includegraphics[width=84mm]{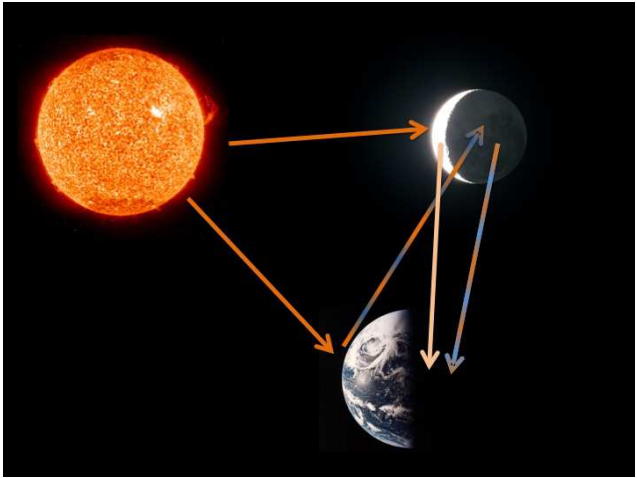}
\caption{Scheme of the path followed by sunlight for both the bight Moon and the earthshine. For the brightside, the light travels directly from the Sun to the Moon, where it is reflected towards the Earth. For the darkside (or earthshine), the light travels from the Sun, to the Earth, it is then reflected towards the Moon, and finally reflected back towards the Earth. The figure is not to scale. \label{fig:earthshine}}
\end{figure}

The earthshine refers to the sunlight reflected from the dayside of the Earth to the dark side of the Moon, and back to the night-side of the Earth. It provides information about the hemispherically-averaged Earth's reflectance (Figure~\ref{fig:earthshine}). The earthshine observation technique requires near simultaneous data from both the Moon's sunlit side, hereafter referred to as the brightside (BS),  and the darker Earth-lit portion of the Moon, hereafter referred to as the darkside (DS). The earthshine provides information about variations in the Earth's albedo, one of the most important parameters to understand the energetic balance of the planet. Along one night, possible variations can be explained by changes in the extent of cloud cover, or by the region of the Earth that is being illuminated during the observations. Continuous observations of this parameter using the earthshine technique makes it possible to characterize yearly and decadal changes in albedo \citep{Qiu03,Palle03}.

The spectral dependence of the Earth's albedo has previously been reported \citep{Woolf02, Arnold02,Montanes04, Hamdani05, Seager05, Turnbull06}. In these studies the earthshine was measured in low spectral resolution to identify the absorption bands present in the Earth's atmosphere, which at visible wavelengths mainly result from water vapour and oxygen \citep{Montanes05}. Additionally, the contribution of vegetation to the Earth's spectrum has also been detected through the slope of the so-called red-edge, the difference in the intensity in the continuum between 680 and 740 nm, due to the reflection at the reddest wavelengths caused by the presence of chlorophyll \citep{Montanes06}.

Going one step further, the earthshine technique applied with high spectral resolution can be used to detect absorption lines from individual atoms present in Earth's atmosphere, specifically from metallic elements, such as Fe, Li, K, Ca or Na. These elements are usually found in atmospheric layers located at altitudes between 75 and 110 km. Their origin is generally assumed to be from the ablation of meteoroids, and they have been regularly observed from the ground using the LIDAR (LIght Detection And Ranging) measurements \citep{Gibson71, Simonich83, Raizada04, Bowman69}. These elements have been used to study the influence of temperature, vertical transport, and meridional wind convergence, in the seasonal structure of the upper atmosphere \citep{Gardner05}. Changes in their density or composition can also provide information about meteor composition after a meteor shower. The signal of an atmospheric metallic component, such as Na {\sc i} doublet, in the earthshine data has been searched for before using the NOT (Nordic Optical Telescope) spectrograph, ALFOSC (Andalucia Faint Object Spectrograph and Camera), with a resolving power of 1000. The results were inconclusive due to a lack of resolving power (\cite{Carrillo11}, private communication).

In the present study, the earthshine technique applied with high-resolution spectroscopy has been used to measure the metallic layers present in the Earth's globally-averaged upper atmosphere. The observations were performed during different nights with the aim of detecting their temporal variability. 

\section{Data}
\begin{figure}
\centering
\includegraphics[width=\columnwidth] {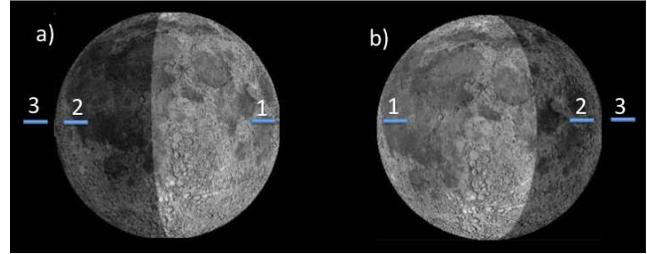}
\caption{Moon phase during the night of a) 2011 February 11 (illumination of 57 per cent.) b) 2011 February 23 (illumination of 60 per cent). The blue lines represent the slit positions during the observations of the brightside (1), the darkside (2) and the sky (3).\label{fig:Lunar_phases}} 
\end{figure}

Observations were taken with the Telescopio Nazionale Galileo (TNG) telescope, at El Roque de Los Muchachos (ORM), using the SARG (Spettrografo Alta Risoluzione Galileo) instrument. SARG is a cross-dispersed echelle spectrograph which covers wavelengths from 370 to 1000 nm. The red ($409-1011$ nm) and blue ($360-514$ nm) cross dispersers used cover the whole available spectral range, and were both set at a resolving power of 29,000, corresponding to a slit size of 0.30$\times$1.15 mm. These data were taken during the nights of 2011 February 11, 12, 23 and 24. The Moon illuminated area was 57 per cent, for the first two nights and 60 per cent, for the last two (lunar phase close to $90^\circ$). For observations at lunar phases closer to New Moon, the observing time is not long enough to get useful data (Moon sets), while at lunar phases closer to Full Moon, the scattered light from the much brighter lunar BS area contaminates the DS images (Figure~\ref{fig:Lunar_phases}). The magnitude of this contamination depends on the phase of the Moon, the airmass and meteorological conditions.

The observations were taken in repetitive cycles: a set of BS spectra, followed by one spectrum of the DS near the lunar edge, then one spectrum of the sky near the lunar edge of the DS selected spot, and finally another set of spectra of the BS (Table~\ref{tbl:Obser_data}). The exposure time of the images changed with airmass along the night, and from one night to the next. Typically, BS exposures were set between 10 and 30 s while DS and sky exposures were set between 900 and 1,200 s. These cyclic observations are the best strategy to correct for the airmass variations from one image to the next. The images were reduced, and the spectra extracted, using the standard software tools from IRAF packages. To obtain more accurate results each of the echelle orders was analysed independently. The wavelength calibration was addressed using ThAr lamps taken the nights of 11th and 24th of February. The dispersion function obtained had an RMS (Root Mean Square) $\sim$0.03 for the red grism and $\sim$0.016 for the blue. It is possible that the pixel-wavelength correlation varies between different nights, however, the Earth's reflection spectrum is calculated through the ratio of images taken consecutively and therefore with the same possible shift. The data reduction also included the removal of cosmic rays using the imlacos software \citep{Van01}. 

\begin{table}
\begin{center}
\begin{tabular}{c c c c c c}
\hline
Date and cycle & BS & DS & Sky & SNR Na & SNR H$\alpha$\\
\hline
11/Feb/2011 \\
Cycle 1 & 13 & 1 & 1 & 270 / 124 / 106 & 520 / 151 / 127\\
Cycle 2 & 21 & 1 & 1 & 360 / 141 / 123 & 505 / 175 / 149\\
Cycle 3 & 21 & 1 & 1 & 440 / 139 / 107 & 600 / 172 / 129\\
\hline
12/Feb/2011 \\
Cycle 1 & 11 & 1 & 1 & 540 / 293 / 255 & 730 / 403 / 362\\
Cycle 2 & 10 & 1 & 1 & 510 / 205 / 119 & 645 / 274 / 153\\
\hline
23/Feb/2011 \\
Cycle 1 &  11 & 1 & 1 & 600 / 98 / 48 & 830 / 129 / 67\\
Cycle 2 &  6 & 1 & 1 & 547 / 96 / 44 & 760 / 125 / 62\\
Cycle 3 & 5 & 1 & 1 & 553 / 90 / 40 & 630 / 118 / 57\\
Cycle 4 & 5 & 1 & 1 & 553 / 93 / 41 & 630 / 120 /57\\
\hline
\end{tabular}
\caption{Number of images taken for each observation night and cycle for BS, DS and Sky. In columns 5 and 6, the mean signal-to-noise ratio (SNR) for each image type (BS/DS/Sky) is given for the spectrum in the Na and H$\alpha$ orders, respectively. \label{tbl:Obser_data}}
\end{center}
\end{table}

When observing the earthshine, it is necessary to make an accurate correction of the background sky, heavily affected by dispersed scattered light from the brightside of the Moon. This effect is very important in the DS images because its surface brightness is small, $\sim$14 \mbox{mag. arcsec\textsuperscript{-2}} \citep{Montanes07}, which means that the sky background makes an important contribution of 73 per cent of the total counts. To minimize this effect, the sky exposure was taken as close in time and airmass as possible to the DS. Finally, to obtain the reflection spectrum of the Earth, the division between the DS spectrum, corrected from the sky contribution, and the BS is calculated \citep{Montanes05}:

\begin{equation}\label{ecuacion}
E_G=\frac{DS-S}{BS}
\end{equation}
 
where $E_{G}$ is the Earth's globally-averaged reflection spectrum, DS is the spectrum from the darkside of the Moon, S is the Sky spectrum, and BS the spectrum from the brightside of the Moon. This procedure results in the removal of the spectral lines that belong to the solar spectrum, the contributions due to reflection from the lunar surface, and the absorption spectrum of the local telluric atmosphere above the observatory. Ideally, one should also remove the sky contribution from the BS images, but this correction is so small ($<$1 per cent) that it only introduces noise. The BS used to calculate the Earth's reflection spectrum for each observing cycle was obtained by averaging the spectra taken at the beginning and at the end for each observing cycle. This calculation provides a final BS spectrum for each cycle at a mean air mass comparable to the DS spectrum's.

The uncertainties in the spectra have been calculated considering that the number of electrons per pixel measured in the CCD, N, follows a Poisson distribution. The error obtained for each pixel of the spectrum is given by $\sqrt{N}$, and also takes account of readout noise. The Earth's reflection spectrum is obtained by indirect methods, operating with the spectra obtained directly from the 2D images. Thus, the error associated with this spectrum should be calculated, using error propagation theory, as

\begin{equation}\label{ecuacion2}
\bigtriangleup f(x_{i})=\sqrt{\sum_{i=1}^{n}{\left( \frac{\partial f }{\partial x_{i}}\bigtriangleup x_{i}\right) }^{2}}
\end{equation}

where \textit{f} is a function depending on the $x_{i}$ variables, $\partial$ represents the partial derivative and $\Delta$ $x_{i}$ the variable uncertainty. More details about the steps followed to calculate the errors are given in Appendix\ref{sec:Apendice}.

\section{Results}
\begin{figure}
\includegraphics[width=\columnwidth] {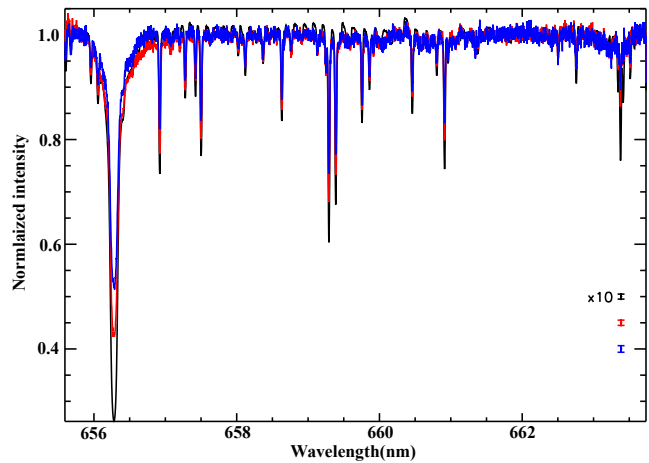}
\caption{Bright side (black), dark side (red) and sky (blue) spectra taken on the 2011 February 11. The figure shows the echelle order containing the $H\alpha$ line (656.27 nm). The spectra are normalized to their pseudo-continuum for comparison. The mean error bar for each spectrum has been plotted in the lower right side of the figure to illustrate the uncertainties. The BS error bar has been multiplied by 10 for comparison. \label{fig:H_BS_ES_S}}
\end{figure}

One of the most characteristic atomic lines in the solar spectrum is the $H\alpha$ line at 656.27 nm. As this line has a solar origin, it should disappear when we calculate the Earth's reflectance spectrum. In Figure~\ref{fig:H_BS_ES_S} the spectrum of 2011 February 11 for the BS, DS and sky in the echelle order in which this line is have been plotted. The actual number of counts per second for the three spectra has a difference of three orders of magnitude. Thus, we have normalized their continuua to allow for a direct comparison. This correction has eliminated the shape of the pseudo-continuum due to the blaze function of the echelle grating, characterized by a stronger sensitivity in the centre of the wavelength window for each order.

\begin{figure}
\includegraphics[width=\columnwidth]{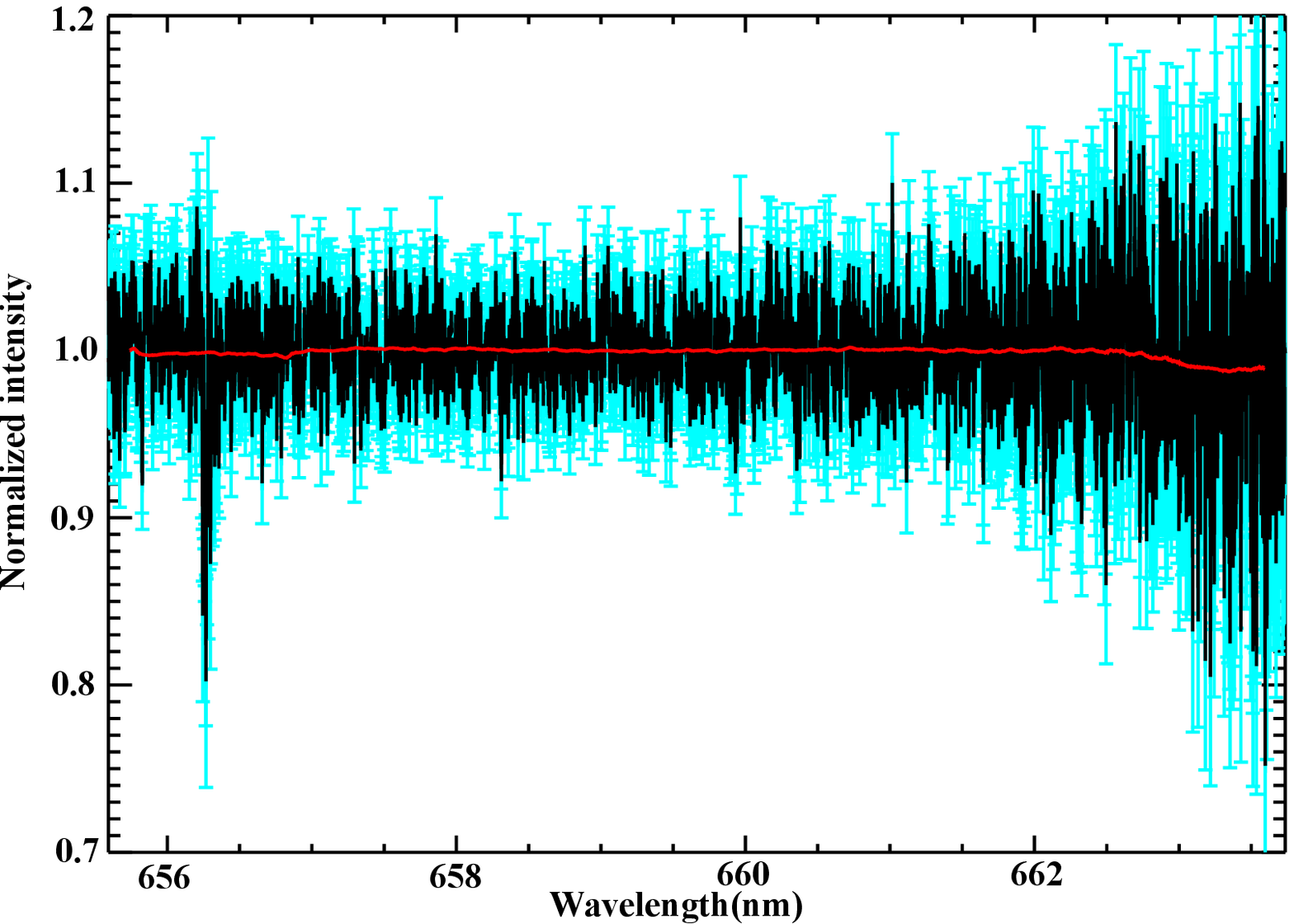}
\includegraphics[width=\columnwidth]{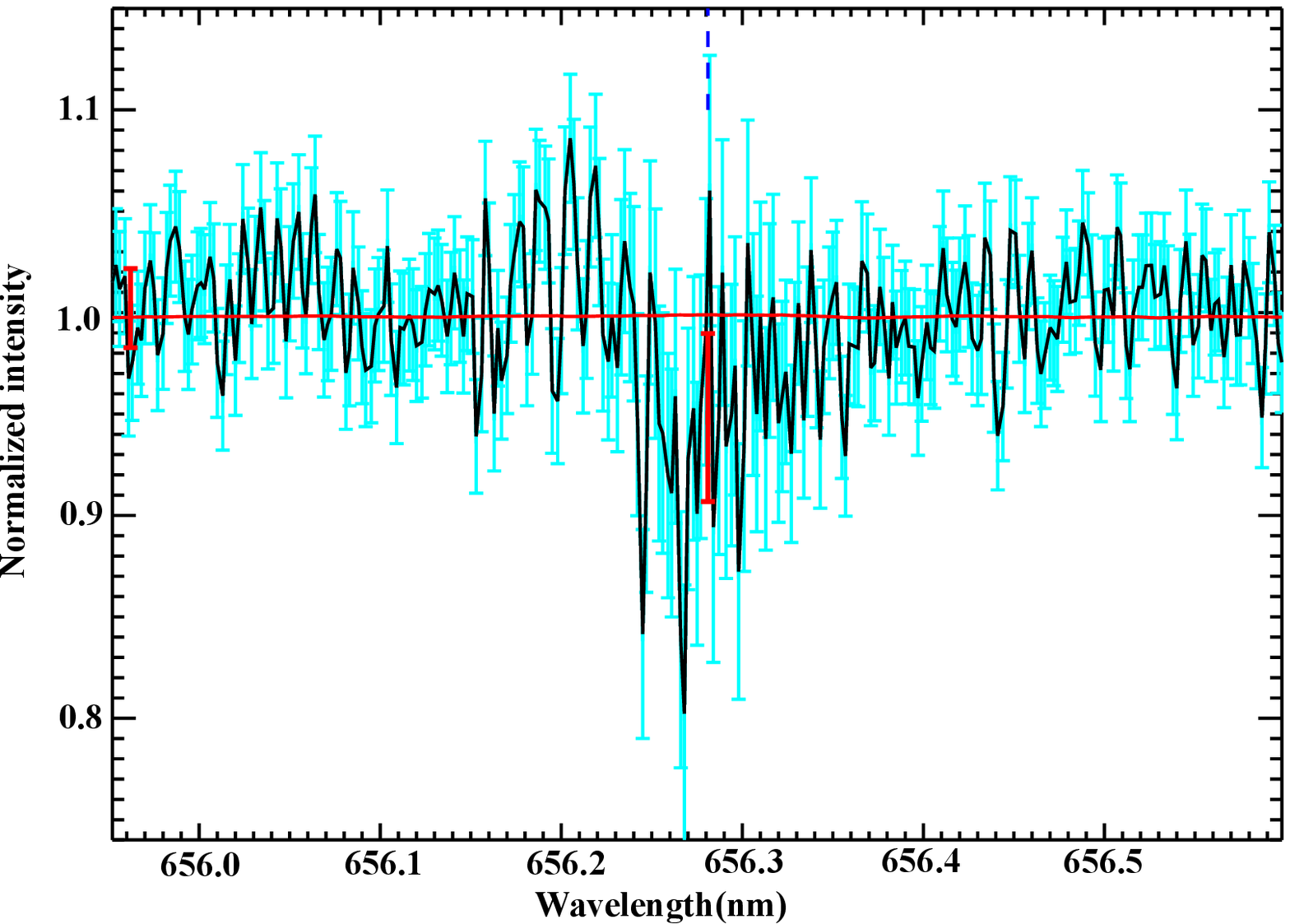}
\caption{Top panel: Earth's reflection spectrum for the $H\alpha$ echelle order with the error bars calculated following the error propagation theory. The red line corresponds to a 500-point running mean. Lower panel: enlargement of the $H\alpha$ residual area. The red error bars show the mean pseudo-continuum position with one standard deviation and the average of the residual, with its uncertainty, obtained with the points inside the half of the FWHM for the BS spectrum. \label{fig:Res_H}} 
\end{figure}

In the top panel of Figure~\ref{fig:Res_H}, the Earth's globally-averaged reflection spectrum obtained using equation~\ref{ecuacion} and the spectra in Figure~\ref{fig:H_BS_ES_S}, is shown for the echelle order containing the $H\alpha$ line. The error bars (light blue) for each point of the spectrum are marked, the solid red line represents the pseudo-continuum and has been calculated with a 500-point running-mean. The noise of the order increases towards the red extreme of the spectrum due to the blaze echelle function diminishing the number of counts. In the lower panel, a zoom of the upper panel centred on the $H\alpha$ line region is plotted. The vertical dashed line marks the position of the $H\alpha$ line centre. A residual signal from the $H\alpha$ line can still be seen. This residual, however, does not have the structure of a real line, with wings, and its deepest point does not match with the centre of the line. It likely remains due to the change in airmass during the observing cycle, which prevented the sky subtraction from being absolutely accurate. Thus, we need to compare the pseudo-continuum and the residual signal to determine if the residual is significant.

The mean value of the pseudo-continuum has been obtained as the average of 400 points around the residual (200 before and 200 after) with the error taken as the standard deviation, $\sigma$, of all the points used in the average. For the residual, the mean value and standard deviation are taken as the average value of the points inside a half of the full-width-at-half-maximum (FWHM) of the $H\alpha$ line in the BS spectrum. The pseudo-continuum value is in normalized intensity units, 1.002$\pm$0.019, and the $H\alpha$ residual 0.9482$\pm$0.041, the centre of the representative point of the residual is then 2.76$\sigma$ away from the continuum.

\begin{figure}
\includegraphics[width=\columnwidth]{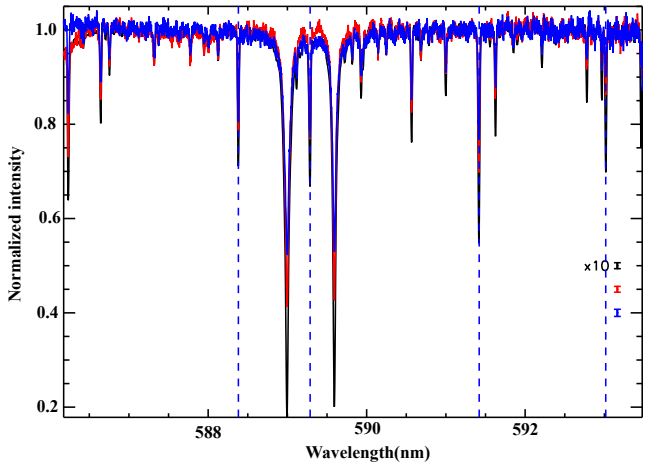}
\caption{As in Figure~\ref{fig:H_BS_ES_S}, but for the echelle order corresponding to the Na doublet at 588.99 nm and 589.59 nm. The vertical dashed lines mark the position of the lines Ni {\sc i} (589.28 nm) and Fe {\sc i} (588.38, 591.42 and 593.01 nm).\label{fig:Na_BS_ES_S}}
\end{figure} 

\begin{figure}
\includegraphics[width=\columnwidth]{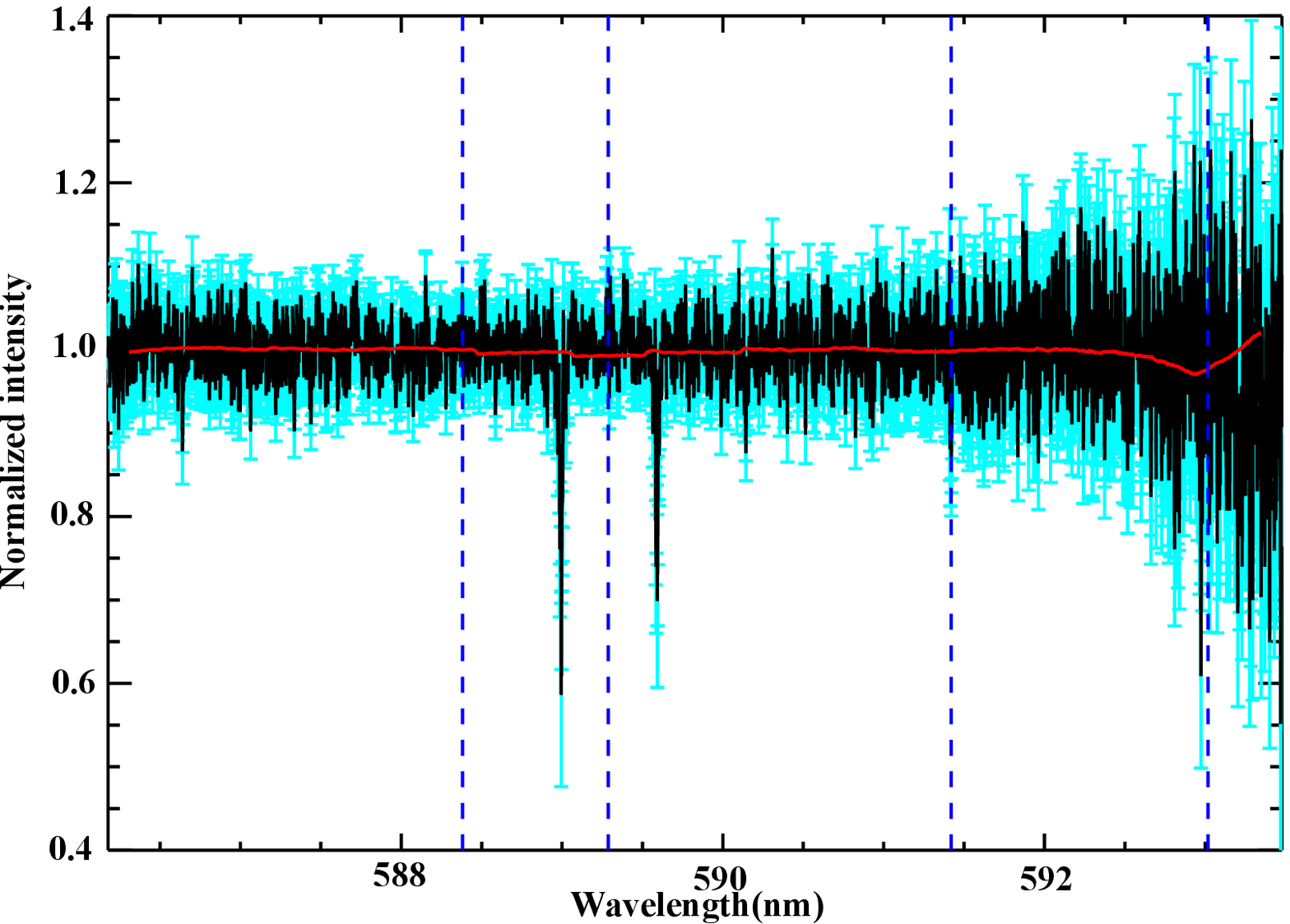}
\includegraphics[width=\columnwidth]{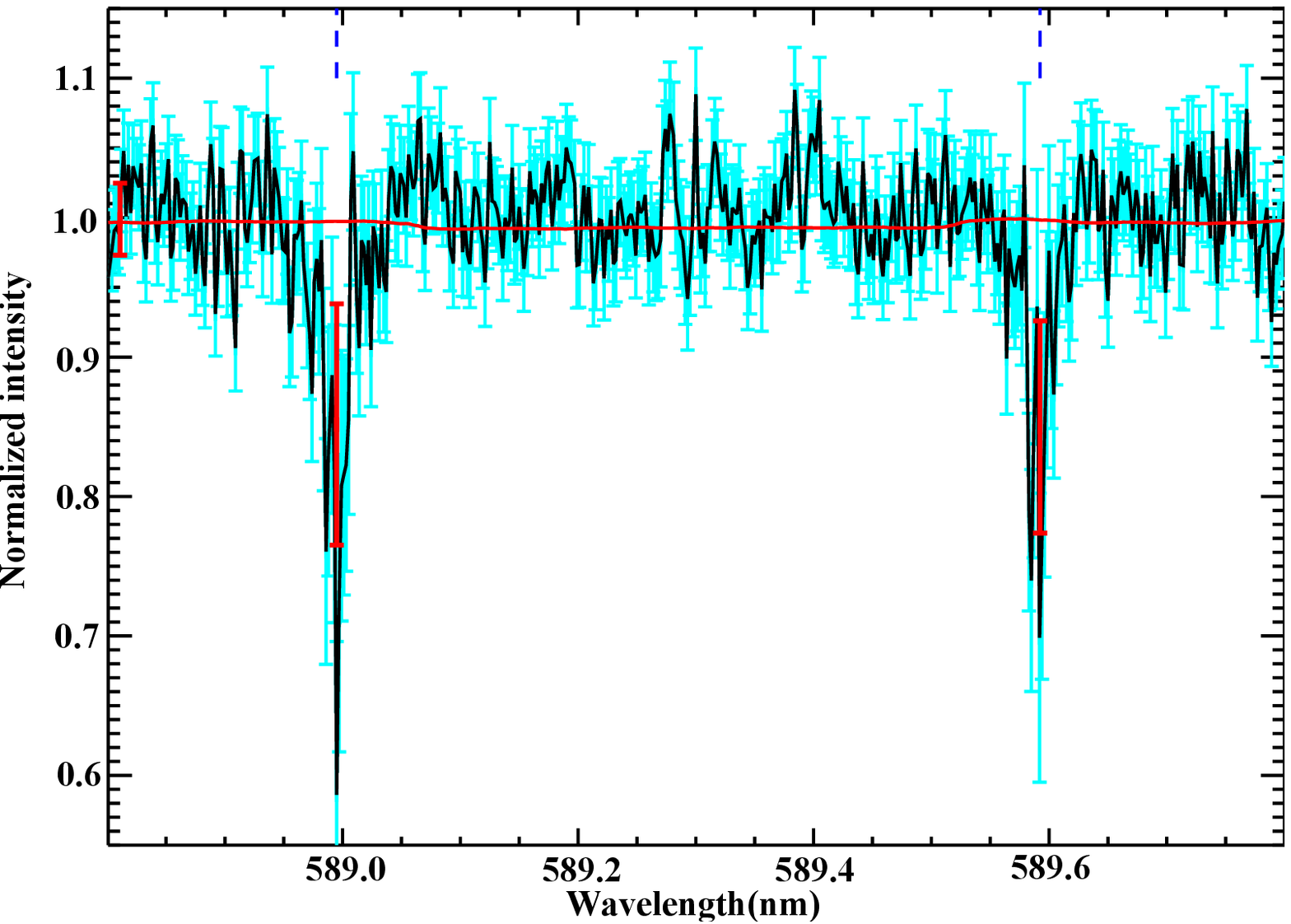}
\caption{As in figure~\ref{fig:Res_H} but for the Na echelle order. The vertical dashed lines in the top panel point out the lines corresponding with Ni {\sc i} (589.28 nm) and Fe {\sc i} (588.28, 591.42 and 593.01 nm) and in the lower panel the Na lines position. \label{fig:Prof_Na}}
\end{figure}

In Figure~\ref{fig:Na_BS_ES_S} the spectra for the night of the 11th of February for the BS, DS and S have been plotted for the echelle order containing the Na doublet (588.9 nm and 589.5 nm). The uncertainties for each pixel have been calculated and represented as in Figure~\ref{fig:H_BS_ES_S}. The Earth's reflection spectrum (Figure~\ref{fig:Prof_Na}) has also been calculated for this order. The dashed vertical lines mark the position of the Ni {\sc i} (589.28 nm) and Fe {\sc i} (588.38, 591.42 and 593.01 nm) atomic lines. These lines are clearly visible in Figure~\ref{fig:Na_BS_ES_S} but they are below the 1$\sigma$ noise level when the Earth's reflection spectrum is calculated, unlike the Na lines which remain the most significant features in this order. The fact that Na lines are still detectable above the noise indicates that its origin is to be found within the Earth's atmosphere. In the lower panel of Figure~\ref{fig:Prof_Na} it can be seen that these lines have the form of proper lines, with well defined wings. Furthermore, due to the high-resolution of the data, the two components of each line are resolved in our spectra.

To address the question of whether these lines are statistically significant, the same method as that applied to the $H\alpha$ residual has been used. The mean value of the pseudo-continuum and its standard deviation have been obtained using 400 points around the Na lines. The Na lines have been represented by the average of the points inside a half of the FWHM in the BS spectrum. In normalized intensity units, the representative value of the continuum is 0.9989, with $\sigma$=0.025, and for the lines the values obtained are 0.851$\pm$0.086 (at 588.9 nm) and 0.849$\pm$0.076 (at 589.5 nm), which correspond to 5.71$\sigma$ and 5.78$\sigma$ detections.

In a Gaussian distribution, 68.27 per cent of data points lie within 1$\sigma$ of the mean value, 95.45 per cent lie within 2$\sigma$, and 99.73 per cent of the values within 3$\sigma$. The two-tailed confidence intervals corresponding to the Na and the H$\alpha$ residual lines are p$<$5.78$\times$10$^{-3}$ and p$<$1.2$\times$10$^{-8}$ respectively. The statistical significance indicates that both the H$\alpha$ residual and Na lines are distinguishable from the propagated noise. However, the lower significance of the deeper absorption feature (H$\alpha$), and the different shape of the residual feature (line profile), point to the difference in their physical origin (bad correction against detection).

During the data analysis procedure we have assumed that the density of the Na layer above the observatory was constant, i.e., that it's contribution is the same in the BS, DS and sky images, and it would disappear when the Earth's reflection spectrum was calculated. However, a change in the density of that layer might have taken place during the observational cycle, increasing the depth of the Na absorption for some of the images. To check this possibility, the depths of both Na lines were calculated for all the BS images taken along the night. If the Na layer doesn't change, only a linear increase due to changing airmass is to be expected. For comparison, the depths of two other solar lines (Fe {\sc i} 588.48 nm and Fe {\sc i} 589.27 nm), which lie in the same echelle order, and are not present in the Earth's atmosphere were also used. Although small deviations from a perfect airmass function are present (due to weather/atmospheric transparency variability), these variations are identical in time for the four lines. Thus, we conclude that possible changes in the Na layer above the telescope are not affecting the detection of the integrated earthshine Na signal.

Another possibility is that the Na detections originate in the Moon's atmosphere (\citet{Flynn93}, and references therein). However, due to the method applied to obtain the Earth's reflection spectrum, this contribution should disappear when the ratio between the DS and BS is calculated. Only if the Moon's atmosphere was denser in the BS area than in the DS area, would any Na originating from the Moon's atmosphere not be corrected for. This is unlikely as the Moon has a Na tail, homogeneously extending the Moon's atmosphere over our sample regions, due to the influence of the solar wind \citep{Smith99}.

Thus, we conclude that the Na lines visible in our Earth's reflection spectrum are originated in the Na layer present in the upper atmosphere in the sunlit hemisphere of the Earth.

\begin{table*}
\begin{center}
\begin{tabular}{c c c c c}
\hline
Date and cycle & Na {\sc i} 588.9 nm ($\sigma$) & Na {\sc i} 589.5 nm ($\sigma$)& Air mass & SNR \\
\hline
11/Feb/2011 \\
Cycle 1 & 5.71 & 5.78 & 1.20 & 19.08\\
Cycle 2 & 9.40 & 6.58 & 1.36 & 19.10\\
Cycle 3 & x & x & 1.69 & 2.77\\
\hline
12/Feb/2011 \\
Cycle 1 & --- & --- & 1.00 & 61.46\\
Cycle 2 & 5.26 & --- & 1.02 & 117.11\\
\hline
23/Feb/2011 \\
Cycle 1 &  --- & --- & 1.96 & 19.36\\
Cycle 2 &  2.10 & --- & 1.73 & 15.44\\
Cycle 3 & --- & 2.06 & 1.64 & 21.39\\
Cycle 4 & 2.54 & --- & 1.60 & 22.16\\
\hline
\end{tabular}
\caption{Intensity differences between the pseudo-continuum and the average point of the lines expressed in pseudo-continuum standard deviation ($\sigma$). Only lines over 2$\sigma$ have been considered. Values over 2$\sigma$ but with a non-accurate sky subtraction, are marked with and x. The mean point of the line was obtained as the average of all the points inside the half of the FWHM, value calculated from the BS spectrums. \label{tbl:Obser_days}}
\end{center}
\end{table*}

The rest of the nights of our observing campaign have been analysed following the same procedure as for the observations on 2011 February 11. The significance of the Na lines for each cycle, expressed as standard deviation of the pseudo-continuum are given in Table~\ref{tbl:Obser_days}. In the Table the values of the SNR of each cycle and the mean air mass of the observations have been included.

Possible differences in the detection between cycles and especially between nights can be related to weather conditions and not only with changes in the airmass. The presence of cirrus cloud or high humidity can lead to a larger scattering from the brightside of the Moon. This effect makes the correction of the sky in the darkside images more difficult. In one case, cycle 3 of 2011 February, 11 the Na detection was found to be over the 3 sigma level, however, other solar lines were also detectable in the earthshine spectrum. Consequently these detections have been associated with an incorrect sky subtraction and thus rejected.

\begin{figure}
\includegraphics [width=\columnwidth]{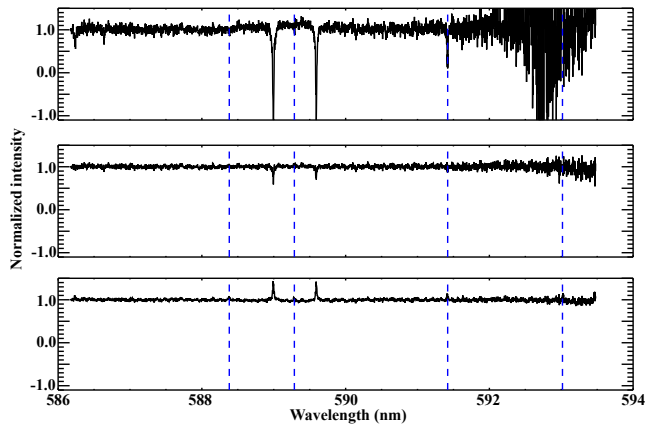}
\caption{Earth's transmission spectrum for the night of 11 of February if a sky correction factor of 1.2 (top panel), 1 (middle panel) and 0.8 (lower panel) is applied. Vertical dashed lines point out solar lines corresponding with Ni {\sc i} (589.28 nm) and Fe {\sc i} (588.28, 591.42 and 593.01 nm). \label{fig:Fact_cielo}}
\end{figure}

To illustrate the effect that an incorrect sky correction can have in the final spectra, in Figure~\ref{fig:Fact_cielo} the Earth's reflection spectrum for the night of the 2011 February 11, calculated following Equation~\ref{ecuacion}, has been represented with three different sky corrections, considering factors of 20 per cent to clearly illustrate the possible effects on the final spectrum. The top panel shows the result of multiplying the sky by 1.2 simulating an overestimated subtraction, the centre panel uses the original sky spectrum, and the lower panel corresponds with a sky multiplied by 0.8, an underestimated correction. The same scale in the \textit{y}-axis has been used in all the panels for comparison. In the top panel the spectral lines become deeper, and in particular solar lines like the one at 593.01 nm can be clearly recognised above the noise. In the bottom panel the under-corrected atmospheric and solar lines become emission lines. These features in each case can be used to identify if the sky correction has been applied properly. A multiplicative factor of 0.94 (6 per cent) is needed to remove the sodium lines. However, this multiplicative factor has to be applied to all orders, and such a correction leads to the appearance of emission lines in alomost all orders, introducing artefacts into the data. Throughout this manuscript, we have not used any multiplicative factor in our analysis.

We have searched for other absorption lines in the rest of the echelle orders, where other atomic lines belonging to the Earth's atmosphere could have been detected. Although there are some hints of detection, we cannot confirm them because although these lines are found at their theoretical wavelengths, more intense lines of the same element are not detected. This could be because they fall in a CCD region with a low signal-to-noise ratio or because they are located inside an atmospheric absorption band. The lines for which we may have tentative detections are: Fe {\sc i} at 372.76, 374.33, 374.94, 375.82, 404.81, 492.05, 496.60, 513.36, 509.07, 594.93 and 602.40 nm, Ca {\sc i} at 422.67, 443.56, 445.47 and 445.58 nm and Na {\sc i} at 568.81 nm. These tentative detections, however, will need further data for confirmation.

\section{Discussion}
The Na layer in the Earth's atmosphere is normally located between 80 and 90 km height, in the mesosphere. The abundance of this metallic ion has a seasonal variation, with a maximum one month before the winter solstice and a minimum close to the summer solstice \citep{Hunten67,Megie78}. Its abundance has been observed to be highly dependent on meteoric activity \citep{Goldberg73}. Using LIDAR measurements, the Sporadic Sodium Layers (SSLs) were discovered in 1978 \citep{Clemesha78}. SSLs are very dense layers, of only a couple of kilometres vertical thickness, which form in a matter of minutes and persist for periods of hours \citep{Senft89}. SSLs form mostly between 15:00 and 00:00 local time, and the presence of the sporadic ionospheric E layer is related to its appearance \citep{Beatty89}.

\begin{figure}
\centering
\includegraphics[width=\columnwidth] {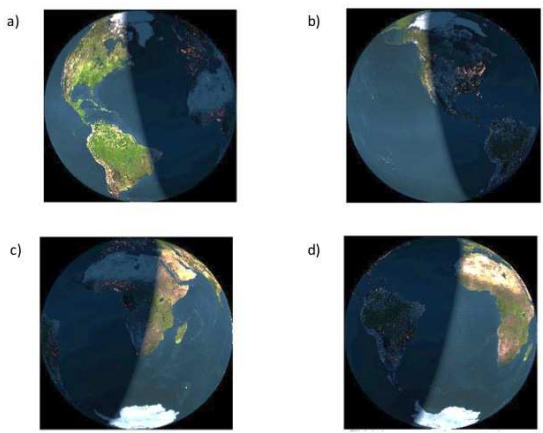}
\caption{The earthshine-contributing areas of the Earth as observed from the Moon for a) 21:00 UT on 11 February 2011, b) 01:00 UT on 12th February 2011, c) 04:00 UT on 24th February 2011 and d ) 07:00 UT on 24th February 2011.}
\label{fig:Tierra}
\end{figure}

In Figure~\ref{fig:Tierra} the region of the Earth visible from the Moon at the beginning and at the end of the observations is shown for two nights for which we have observations. The area illuminated by the Sun changes from the Canary Island's westwards view (America and the Pacific ocean) in the first two nights (11 and 12 February) to the east view (Africa, Antarctica, Atlantic and Indian oceans) for the las two (23 and 24 of February). The observations include very different regions of the world, but the Na lines are only visible in some cycles, independently of the geographic area observed. This could be related with the presence of SSLs in the upper atmosphere. Due to the small SNR we have in our data, SNR$\sim$20, it is possible that a Na signal was visible only in the cycles in which a SSLs existed. The fact that these episodes last only a few hours could explain why the Na lines are not visible in all the cycles taken.

Previous studies of the Na layers used the LIDAR technique during specific observation campaigns, getting local atmosphere information during certain time periods. Since the earthshine technique provides integrated data from the whole Earth's dayside visible area from the Moon, LIDAR data are not useful to validate our results. The best source of information would be Earth's Na dayside satellite observations, but unfortunately, such data do not exist. 

It is also worth mentioning that the H$\alpha$ absorption line is present in the solar spectrum but not in the Earth's atmosphere. However this line can be found in emission in Earth's spectrum and has its origin in the helium present in the geocorona. This H$\alpha$ emission is a result of a fluorescence process initiated by Lymman $\alpha$ (121 nm) solar photons \citep{Sahan07}, and it is used to chart the geocorona as its illumination depends on the time and the geometry of the observations \citep{Shih85,Nossal93,Nossal97,Nossal98}. However, the H$\alpha$ residuals in our spectra are not related with the emission from the Earth's geocorona, but with the incomplete sky subtraction to the darkside. If this correction was perfect, and supposing the geocorona was detectable, such detection would appear as an emission line in our spectra.

\section{Conclusions}
The Earth's reflectance spectrum has been observed with the earthshine technique using SARG at the TNG telescope, with a resolving power of $R=29,000$. This experiment is the first time in which this technique has been applied using such high resolution. This high resolving power has allowed us to test the detectability of atomic lines in the Earth's averaged reflectance spectrum. The most difficult part of the analysis of these dataset is the precise correction for the background sky. Nevertheless, by imposing high confidence constrains for a detection, we find that atomic Na lines were detected during different observational nights, but not all. We conclude that the origin of these lines lies in the Earth's upper atmosphere, and thus have geological or meteoritic origins. The fact that these lines are not visible along all the nights and for all the observational cycles could be related to the presence of the SSL layer in the Sun-illuminated fraction of the Earth. The known transitional nature of these events could explain why sometimes the Na is not visible in our spectra, while it is detectable a few hours later. Thus, we suggest that rather more continuous observations of the earthshine could be used to monitor and understand the nature and variability of the Na layer at global scales.

Other possible atomic lines of terrestrial origin have been tentatively detected, but they cannot be confirmed with the present data due to the low SNR (which results from such high spectral resolution and the faintest of the earthshine). Confirmation of these possible atomic lines in the future requires observations of the earthshine at an intermediate resolution (around $R=5,000-10,000$), enough to increase the SNR of the observations while keeping the possibility of detecting atomic lines.

\section*{Acknowledgements}
This work was based on observations made with the Telescopio Nazionale Galileo (TNG) operated on the island of La Palma by the Fundaci\'on Galileo Galilei - INAF, Fundaci\'on Canaria in the Spanish Observatorio de Roque de los Muchachos of the Ins\-ti\-tu\-to de As\-tro\-f\'{\i}\-si\-ca de Canarias, and supported by the non-longer existing Spanish Ministry of Ciencia e Innovaci\'on under the grant CGL2009-10641. Beatriz Gonz\'alez-Merino would like to thank Dr. Benjamin A. Laken (IAC) for valuable comments.

\footnotesize{
\bibliographystyle{mn2e}
\bibliography{/Users/bgmerino/Beatriz/IAC/Referencias_Articulos/Articulos}
}

\label{lastpage}

\appendix
\section{Calculus of the spectra error}\label{sec:Apendice}

The calculation of the error in the Earth's transmission spectra has been processed by the following procedure.

First, we quantified the error in the measurement of each of the brightside, earthshine and sky spectra. These errors were obtained assuming that the number of electrons measured, N, by the CCD follows a Poisson distribution. Given this, the uncertainty for each pixel can be obtained as $\sqrt{N}$. This value is given by the apall task in iraf during the extraction of the spectra from the 2D images. This task also takes into account the readout noise of the CCD. These images are the only direct measurements taken during the observations, the rest of the steps to obtain the final Earth's spectra consisted of different operations using these original spectra. To consider the evolution of the errors through all of the data reduction steps, error propagation theory was applied.

Supposing a magnitude, \textit{f}, depending on several variables, (x$_{1}$, x$_{2}$, $\ldots$, x$_{n}$) and is obtained through indirect methods, then its error, \textit{$\Delta$f}, calculated using error propagation theory is given by:
 
\begin{equation}\label{ecuacion3}
\begin{split}
\Delta f & =\sqrt{\left ( \frac{\partial f}{\partial x_{1}} \Delta x_{1}\right )^{2}+\left ( \frac{\partial f}{\partial x_{2}} \Delta x_{2}\right)^{2}+...+\left ( \frac{\partial f}{\partial x_{n}} \Delta x_{n}\right )^{2}} \\
         & = \sqrt{\sum_{i=1}^{n}\left (\frac{\partial f}{\partial x_{i}} \Delta x_{i}\right )^{2}}
\end{split}
\end{equation}

where $\Delta$x$_{1}$, $\Delta$x$_{2}$, $\ldots$, $\Delta$x$_{n}$ represent the uncertainties associated with each variable and $\partial$\textit{f}/$\partial$x$_{n}$ is the partial derivative of the magnitude $f$ with respect of x$_{n}$ variable.

From this, the errors during the calculation of the Earth's transmission spectra for the different steps followed in the data reduction process were assessed; these are detailed below. To clarify the mathematical notation, the value in each of the pixels of the spectrum obtained from the apall task is given by N$\pm$ $\Delta$N=N$\pm$ $\sqrt{N}$, interchanging the letter N for BS in the case of a Brightside image, DS for a darkside and S for a sky. 

Firstly, each spectrum was normalized by its exposure time.   

\begin{equation}\label{ecuacion4}
\Delta BS_{t}=\frac{\Delta BS}{t} \qquad \Delta DS_{t}=\frac{\Delta DS}{t} \qquad \Delta S_{t}=\frac{\Delta S}{t} 
\end{equation}

These are the errors used in the next steps, but with the aim of simplifying the expressions, the t subscript is going to be removed.

The image of the brightside used in the calculation of the Earth's spectrum was obtained by averaging the images taking at the beginning and at the end of the cycle (BS$_{1}$,BS$_{2}$, $\ldots$,BS$_{n}$), and its error is then: 

\begin{equation}\label{ecuacion5}
\Delta BS=\sqrt{\left(\frac{\Delta BS_{1}}{n} \right)^{2}+\left(\frac{\Delta BS_{2}}{n} \right)^{2}+...+\left(\frac{\Delta BS_{n}}{n} \right)^{2}}
\end{equation}

where $n$ is the number of BS used in the average.  

Finally the error for the spectra of the Earth's transmission spectrum is given by:
\begin{equation}\label{ecuacion6}
\Delta (DS-S)=\sqrt{\left(\Delta DS\right)^{2}+\left(\Delta S\right)^{2}}
\end{equation}

\begin{equation}\label{ecuacion7}
\Delta \left(\frac{DS-S}{BS} \right)=\sqrt{\left(\frac{\Delta (DS-S)}{BS} \right)^{2}+\left(\frac{DS-S}{BS}^{2} \Delta BS \right)^{2}}
\end{equation}

\end{document}